\documentclass[review,12pt,times]{elsarticle}
\usepackage[%
  pdftitle={Disturbance Observer based Control of Integrating Processes with Dead-Time using PD controller},%
  pdfauthor={Kadam, Sujay Dilip},%
  bookmarks=true,%
  breaklinks=true,%
  colorlinks=true,%
  linkcolor=blue,anchorcolor=blue,%
  citecolor=orange!80!black,filecolor=blue,%
  menucolor=blue,pagecolor=blue,%
  urlcolor=blue]{hyperref}
  \usepackage[top=0.75in,bottom=0.75in,left=0.75in,right=0.75in]{geometry}

\makeatletter
\usepackage{geometry}
\usepackage{natbib}
\usepackage{amsmath}
\usepackage{amssymb}
\usepackage{caption}
\usepackage{graphicx}
\makeatletter
\def\ps@pprintTitle{%
 \let\@oddhead\@empty
 \let\@evenhead\@empty
 \def\@oddfoot{}%
 \let\@evenfoot\@oddfoot}
\makeatother
\begin{document}

\begin{frontmatter}

\title{Disturbance Observer based Control of Integrating Processes with Dead-Time using PD controller}

\author{Sujay D. Kadam\corref{sujay}}
\ead{kadam.sujay@iitgn.ac.in}
\address{SysIDEA Lab,~IIT Gandhinagar,~India.}
\begin{abstract}
The work presented here addresses the issue of tuning PD controllers for controlling integrating plus dead-time (IPDT) processes using settling time and gain and phase margin specifications. Tuning formulae are derived for PD controller being used for controlling IPDT processes. Simulations have been carried out to determine the effectiveness of the tuning formulae. Also, comparison between proposed and the other existing tuning methods with respect to performance indices, time domain specifications and the control effort expressed in terms of control signal energy has been done. Effects of gain margin and settling time specifications on process output responses have been discussed.  A PD controller cannot account for load disturbances. So, a disturbance observer is suggested for use along with the PD controller. Regulatory and tracking responses are then compared with respect to performance indices and control effort. 
\end{abstract}


\end{frontmatter}

\section{Introduction}
The problem of controlling integrating plus dead time (IPDT) processes has received significant attention during recent past \cite{mercader2017pi,kadam2013control,ali,zhang,wang2}. Numerous PI, PD and PID controller tuning methods for intergrating processes have been proposed earlier \cite{sree,chidambaram,zhong,matausek,luyben,pomerleau}. 
Integrating processes are open loop unstable because of pole at origin. In addition, existence of dead-time along with the integrator makes it difficult to tune a PI, PD or a PID controller for controlling the resulting integrating plus dead-time (IPDT) process. The paper discusses the use of PD control for IPDT process, assuming that the IPDT process, by its integrating nature, should not require integral control action at all. However, to subvert the effects of load disturbances, integral action is required. Instead of using a full three mode PID controller, the use of disturbance observer is suggested for taking into account load disturbances, as integral action necessitates more control effort by the final control element. Simulations have been carried out for ascertaining the performance of the controllers and the results have been compared with methods proposed by Wang and Cluett \cite{wang2}, Sree and Chidambaram \cite{sree} and Ali and Majhi \cite{ali}.

The paper is organized in the following manner. Section 2.1 deals with the derivation of the tuning rules for PD controller using settling time ($T_s$), gain margin ($A_m$) and phase margin ($\phi_m$) specifications. Section 2.2 enumerates a step by step procedure to get tuning parameters for a PD controller to control a given IPDT process. Section 2.3 gives the guidelines for selecting $\omega_{pc}$ and $\omega_{gc}$ frequencies required for designing the PD controller. Section 3.1 starts with assumption of the IPDT process to be controlled  and then controller tuning parameters for methods under comparison are deduced. Comparison of step responses of proposed PD tuning and other methods has been done with respect to time domain specifications like rise time, settling time and overshoot and perfomance indices like ISE, IAE and ITAE in Section 3.2 Also, in Section 3.2, control effort defined in terms of control signal energy required for each method to control the process . Servo responses for proposed and tuning method are compared in Section 3.3. Section 3.4 demonstrates how a change in either settling time ($T_s$) or gain margin ($A_m$) specification causes a change in process output response at a constant value of phase margin. The idea of using disturbance observer alongwith the proposed PD controller for regulatory control is presented in Section 4.1. Regulatory responses for each of the methods under comparison under the influence of unit step disturbance have been compared in Section 4.2 against performance indices and the criterion of control signal energy. Operation of controllers being compared under servo plus regulatory control is discussed in Section 4.3.  Conclusions thus drawn, have been put forth in Section 5. Section 6 includes the references.

\section{Controller Design }
\subsection{Derivation of tuning rules for PD controller}
Consider an integrating plus dead-time (IPDT) process, given by the transfer function
\begin{equation}\label{eq:IPDT}
G_I(s)=\frac{K_p}{s}e^{-ds},
\end{equation}
which is to be controlled by a PD controller of the form,
\begin{equation}\label{eq:Gpd}
G_{C_{PD}}(s)=K_{c_{PD}}(1+T_{d_{PD}}s).
\end{equation}
The loop transfer function is given by,
\begin{equation}\label{eq:Gol}
G_{OL_{PD}}(s)=\frac{K_pK_c(1+T_{d_{PD}}s)}{s}e^{-ds},
\end{equation}
where $K_p$ is the process gain, $d$ is the dead-time, $K_c$ is the proportional gain and $T_d$ is the derivative time.

From the definitions of gain margin,
\begin{equation}\label{eq:Am}
A_m=\frac{1}{|G_{C}(j\omega_{pc})G_{I}(j\omega_{pc})|}
\end{equation}

and from definitions of phase margin,
\begin{equation}\label{eq:phim}
\phi_m={\rm arg}[G_{C}(j\omega_{gc})G_{I}(j\omega_{gc})]+\pi
\end{equation}
where $A_m$ is the gain margin, $\phi_m$ is the phase margin, $\omega_{gc}$ is the gain crossover frequency and $\omega_{pc}$ is the phase crossover frequency.

In context of equation (\ref{eq:phim}), the phase margin for loop transfer function for process with PD controller is given by
\begin{equation}
\phi_{m_{PD}}={\rm arg}[G_{OL_{PD}}(j\omega_{gc_{PD}})]+\pi
\end{equation}
and
\begin{equation}
{\rm arg}[G_{OL_{PD}}(j\omega_{gc_{PD}})]=\arctan(\omega_{gc_{PD}}T_{d_{PD}})-\omega_{gc_{PD}}d-\frac{\pi}{2}.
\end{equation}
We get the following equation on simplification
\begin{equation}
\tan(\phi_{m_{PD}}+\omega_{gc_{PD}}d-\frac{\pi}{2})=\omega_{gc_{PD}}T_{d_{PD}}
\end{equation}
Re-arranging the equation, we have,
\begin{equation}\label{eq:Td}
T_{d_{PD}}=\frac{\tan(\phi_{m_{PD}}+\omega_{gc_{PD}}d-\frac{\pi}{2})}{\omega_{gc_{PD}}}
\end{equation}

where, $\phi_{m_{PD}}$ is the phase margin and $\omega_{gc_{PD}}$ is the gain crossover frequency for loop transfer function defined by equation ({\ref{eq:Gol}}). The value of $\omega_{gc_{PD}}$ has to be pre-specified so as to achieve desired closed loop performance.
\\
\\
Also, in context of equation ({\ref{eq:Am}}), we can write the following equation,
\begin{equation}
A_{m_{PD}}=\frac{1}{|G_{C_{PD}}(j\omega_{pc_{PD}})G_{I}(j\omega_{pc_{PD}})|}
\end{equation}
\begin{equation}
\therefore A_{m_{PD}}=\frac{\omega_{pc_{PD}}}{K_p.K_{c_{PD}}\sqrt{1+(T_{d_{PD}}.\omega_{pc_{PD}})^2}}
\end{equation}
The above equation can be re-arranged as,
\begin{equation}\label{eq:Kc}
K_{c_{PD}}=\frac{\omega_{pc_{PD}}}{K_p.A_{m_{PD}}\sqrt{1+(T_{d_{PD}}.\omega_{pc_{PD}})^2}}
\end{equation}
where, $A_{m_{PD}}$ is the gain margin and $\omega_{pc_{PD}}$ is the phase crossover frequency for loop transfer function defined by equation (\ref{eq:Gol}).\\

Equations (\ref{eq:Td}) and (\ref{eq:Kc}) give the derivative time and proportional gain respectively for a PD controller used for controlling an IPDT process given by equation ({\ref{eq:IPDT}).
\\
\subsection{PD controller design procedure}
For the integrating plus dead time process $G_I(s)$ having model as equation ({\ref{eq:IPDT}),
\begin{enumerate}
\item Specify the desired values of gain margin ($A_{m_{PD}}$) and phase margin ($\phi_{m_{PD}}$).
\item Specify the gain and phase crossover frequencies ($\omega_{gc_{PD}}$) and ($\omega_{pc_{PD}}$) indirectly by specifying desired settling time ($T_{s_{PD}}$) as per equation (\ref{eq:omega}). Refer to Section 2.3 for guidelines on selecting $\omega_{pc}$ and $\omega_{gc}$ frequencies.
\item Calculate $T_{d_{PD}}$ using the values specified above in equation (\ref{eq:Td}). Ignore the negative sign if the value of $T_d$ is negative, as derivative time cannot be negative.
\item The value of $K_{c_{PD}}$ can be obtained by substituting the values of $T_{d_{PD}}$ obtained in the previous step, $A_{m_{PD}}$, $\phi_{m_{PD}}$, and $\omega_{pc_{PD}}$ in equation (\ref{eq:Kc}).
\end{enumerate}

\subsection{Guidelines for selecting $\omega_{pc}$ and $\omega_{gc}$ frequencies}
Performance specifications in terms of frequency domain parameters are not as intuitive as time domain specifications for an average plant operator. The choice of the crossover frequencies $\omega_{pc}$ and  $\omega_{gc}$ for getting reasonable controller parameters can be challenging.

The frequencies $\omega_{pc}$ and  $\omega_{gc}$ can be specified indirectly in terms of settling time $(T_s)$ as described in \cite{wang1}. The desired settling time can be intuitively specified, to find the values of $\omega_{pc}$ and  $\omega_{gc}$ as,
\begin{equation}\label{eq:omega}
\begin{split}
\omega_{pc}&=\frac{2\pi}{T_s}\\
\omega_{gc}&=\frac{4\pi}{T_s}\\
\therefore \omega_{gc}&=2\omega_{pc}
\end{split}
\end{equation}
The frequencies thus selected for computation of controller parameters are the frequencies that have greatest influence on the time domain performance \cite{wang1}.

\section{Simulations and comparison of results with other tuning methods}
\subsection{Consideration of process and controller tuning parameters for simulation}

Simulations are carried out on process defined by equation (\ref{eq:IPDT}) with $K_p=0.0506$ and $d=6~s$.
\begin{equation}\label{eq:IPDTsim}
G_I(s)=\frac{0.0506}{s}e^{-6s}
\end{equation}

The tuning parameters in Wang-Cluett's method \cite{wang2} are selected as $\zeta=1$ and $\beta=3$. The settings for proposed PD controller are obtained by assuming settling time $(T_{s_{PD}}=40~{\rm s})$, gain margin $(A_{m_{PD}}= 2 = 6.0206~{\rm dB})$ and phase margin $(\phi_{m_{PD}}=~\pi~{\rm rad}~=~180  ^\circ$). 
Table \ref{TableControlParameters} shows the controller parameters used in the simulation studies for various tuning methods.
\begin{table}[!htbp]
\caption{Controller Parameters}
\centering
\label{TableControlParameters}
\begin{tabular}{ c  c  c  c }
\hline
\hline
Parameters &$K_c$ &$T_i$ & $T_d$ \\
& &(sec) &(sec) \\
\hline
\hline
Wang-Cluett's tuning &1.2416 &55.065 &1.028\\
\hline
Sree-Chidambaram's &2.95 &15 &3\\
tuning &  &  & \\
\hline
Ali-Majhi's tuning &3.39 &19.02 &2.94\\
\hline
Proposed PD tuning &1.5321 &  &1.0343\\
\hline
\hline
\end{tabular}
\end{table}

\subsection{Comparison of tuning methods with respect to step responses}
The step responses for IPDT process given in (\ref{eq:IPDTsim}) for respective methods are shown in Fig. \ref{fig1}. The set point used is a unit step signal.

\begin{figure}[!htbp]
\centering
\includegraphics[width=0.6\linewidth]{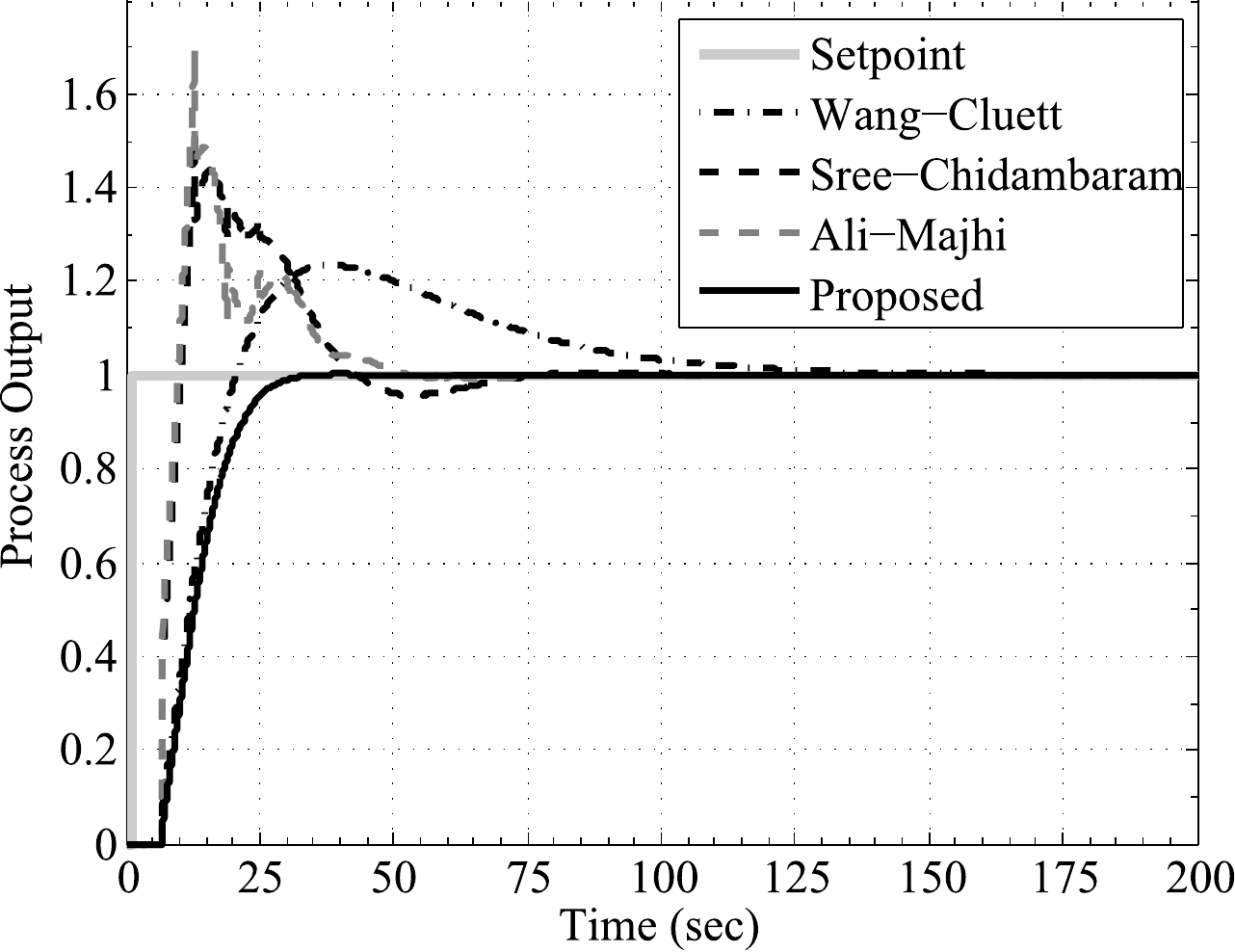}
\caption{Tracking responses for unit step reference command}
\label{fig1}
\end{figure}

The comparison of the responses is made based on the time domain specifications of rise time, settling time and overshoot. This comparison is tabulated in Table \ref{table1}.

\begin{table}[!htbp]
\caption{Unit step reference command: Comparison of time domain specifications}
\centering
\label{table1}
\begin{tabular}{ c  c  c  c }
\hline
\hline
Parameters &Rise &Settling &Overshoot \\
&Time &Time & \\
&(sec) &(sec) &(\%) \\
\hline
\hline
Wang-Cluett's tuning &10.65 &110.97 &23.43\\
\hline
Sree-Chidambaram's &2.60 &65.75 &52.24\\
tuning &  &  & \\
\hline
Ali-Majhi's tuning &2.03 &47.91 &69.56\\
\hline
Proposed PD tuning &14.03 &28.12 &0.19\\
\hline
\hline
\end{tabular}
\end{table}
From the comparisons made from Table \ref{table1}, it can be inferred that proposed PD tuning method provides smaller values of settling time and overshoot.

Table \ref{table2} shows comparisons with respect to performance indices like IAE, ITAE and ISE when the setpoint is a step signal.
\begin{table}[!htbp]
\caption{Comparison of step responses with respect to Performance Indices}
\centering
\label{table2}
\begin {tabular}{ c  c  c  c }
\hline
\hline
Performance & & &\\ 
Indices &ISE &IAE &ITAE\\
\hline
\hline
Wang-Cluett's tuning &12.27 &26.81 &1067\\
\hline
Sree-Chidambaram's tuning &8.774 &15.55 &243.2\\
\hline
Ali-Majhi's tuning &8.264 &13.61 &173\\
\hline
Proposed PD tuning &0.2512 &1.022 &4.479\\
\hline
\hline
\end{tabular}
\end{table}
From Table \ref{table2}, proposed tuning gives rise to smaller values of IAE and ITAE indices meaning that it provides smaller values of overshoot and settling time.

Another criterion considered for comparison is the control effort defined in terms of the energy spent by the controller in controlling the process i.e. the energy of the control signal $(E_{u(t)})$ defined by,
\begin{equation}\label{eq:ControlEnergy}
E_{u(t)}= \int_{0}^{T}{|u(t)|^2}dt
\end{equation}
where, $t$ is the simulation time and $[0~T]$ is the simulation time interval. Comparison results are tabulated in Table \ref{table3}.

\begin{table}[!htbp]
\caption{Comparison with respect to control signal energy as defined in (\ref{eq:ControlEnergy}) for simulation interval [0 200].}
\centering
\label{table3}
\begin {tabular}{ c  c }
\hline
\hline
Parameter &Control signal energy\\
\hline
\hline
Wang-Cluett's tuning &214.9\\
\hline
Sree-Chidambaram's tuning &4351\\
\hline
Ali-Majhi's tuning &5690\\
\hline
Proposed PD tuning &78.22\\
\hline
\hline
\end{tabular}
\end{table}

\subsection{Comparison of tuning methods with respect to servo responses}
In order to judge the proposed tuning methods set point tracking ability, the set points were applied to the closed loop system as,
 \[
    {\rm SP} =\left\lbrace \begin{array}{lr}
            1, & \text{for } t \in [0 ~100]\\
        3, & \text{for } t \in (100~200]\\
        2, & \text{for } t\in (200~300].
        \end{array}\right.
  \]
Fig. \ref{fig2} shows the set point tracking responses.
\begin{figure}[!htbp]
\centering
\includegraphics[width=0.6\linewidth]{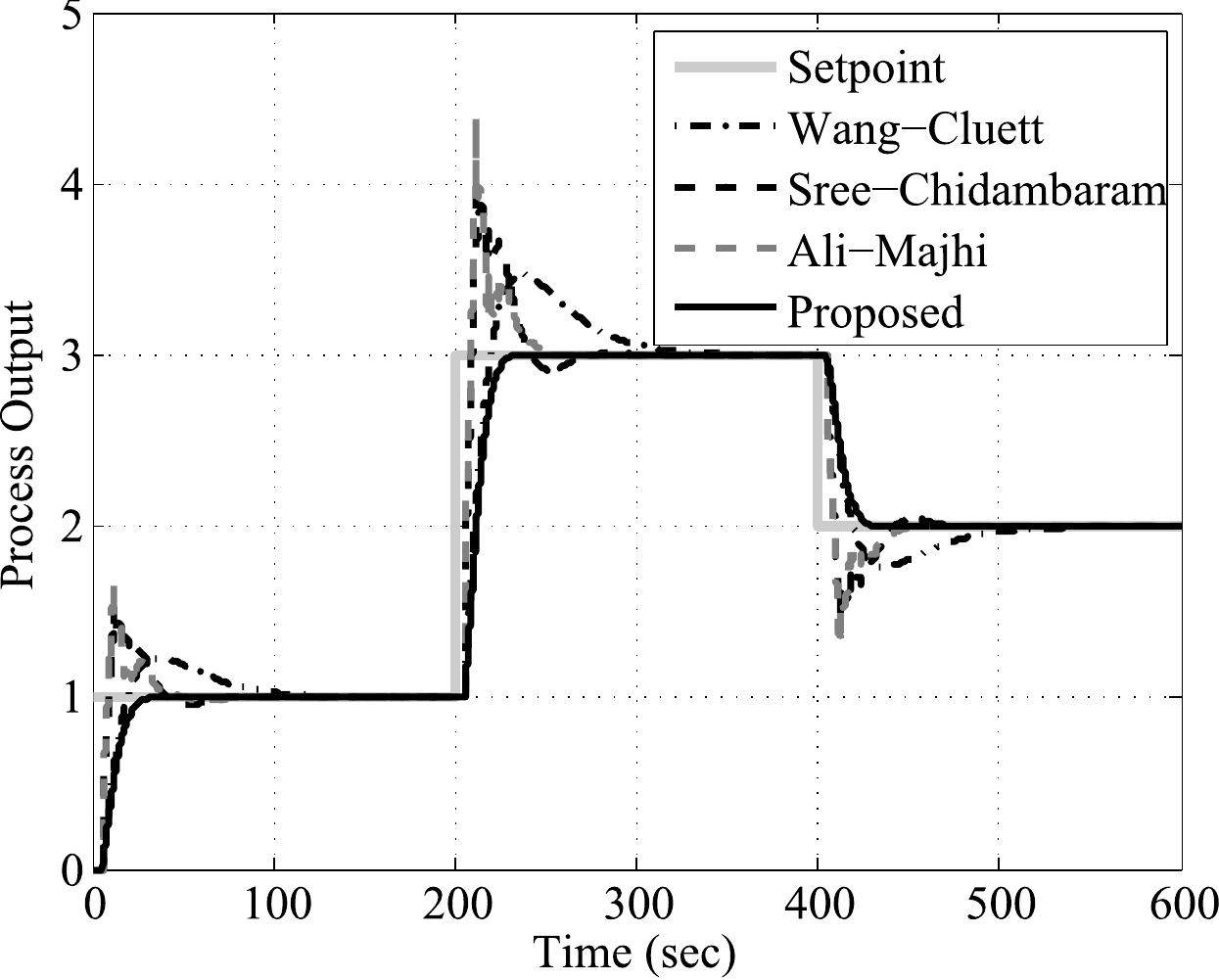}
\caption{Servo responses of the tuning methods being compared}
\label{fig2}
\end{figure}
The performance indices IAE and ISE are compared for each set point change in following Tables \ref{tab_servo_ISE} and \ref{tab_servo_IAE} respectively.
\begin{table}[!htbp]
\caption{Comparison of Servo responses against ISE criterion}
\centering
\label{tab_servo_ISE}
\begin{tabular}{ c  c  c  c }
\hline
\hline
Parameters &\multicolumn{3}{c}{ISE} \\ \cline{2-4}
 &SP=1 &SP=3 &SP=2\\
\hline
\hline
Wang-Cluett's tuning &10.91 &43.63 &10.92\\
\hline
Sree-Chidambaram's tuning &8.774 &35.1 &8.779\\
\hline
Ali-Majhi's tuning &8.264 &33.05 &8.256\\
\hline
Proposed PD tuning &9.887 &39.55 &9.887\\
\hline
\hline
\end{tabular}
\end{table}

\begin{table}[!htbp]
\caption{Comparison of Servo responses against IAE criterion}
\label{tab_servo_IAE}
\centering
\begin{tabular}{ c  c  c  c }
\hline
\hline
Parameters &\multicolumn{3}{c}{IAE} \\ \cline{2-4}
 &SP=1 &SP=3 &SP=2\\
\hline
\hline
Wang-Cluett's tuning &22.7 &45.39 &22.7\\
\hline
Sree-Chidambaram's tuning &15.56 &31.11 &15.58\\
\hline
Ali-Majhi's tuning &13.63 &27.26 &13.64\\
\hline
Proposed PD tuning &12.95 &25.9 &12.95\\
\hline
\hline
\end{tabular}
\end{table}

\subsection{Comparison of process output responses for different values of gain margin $(A_m)$ and settling time $(T_s)$ specifications}
The effect of input specifications like gain margin $(A_m)$ and settling time $(T_s)$ as described in Section 2.2, on the response of a process given by equation (\ref{eq:IPDT}) controlled by the resulting PD controller is demonstrated here.
\begin{figure}[!htbp]
\centering
\includegraphics[width=0.6\linewidth]{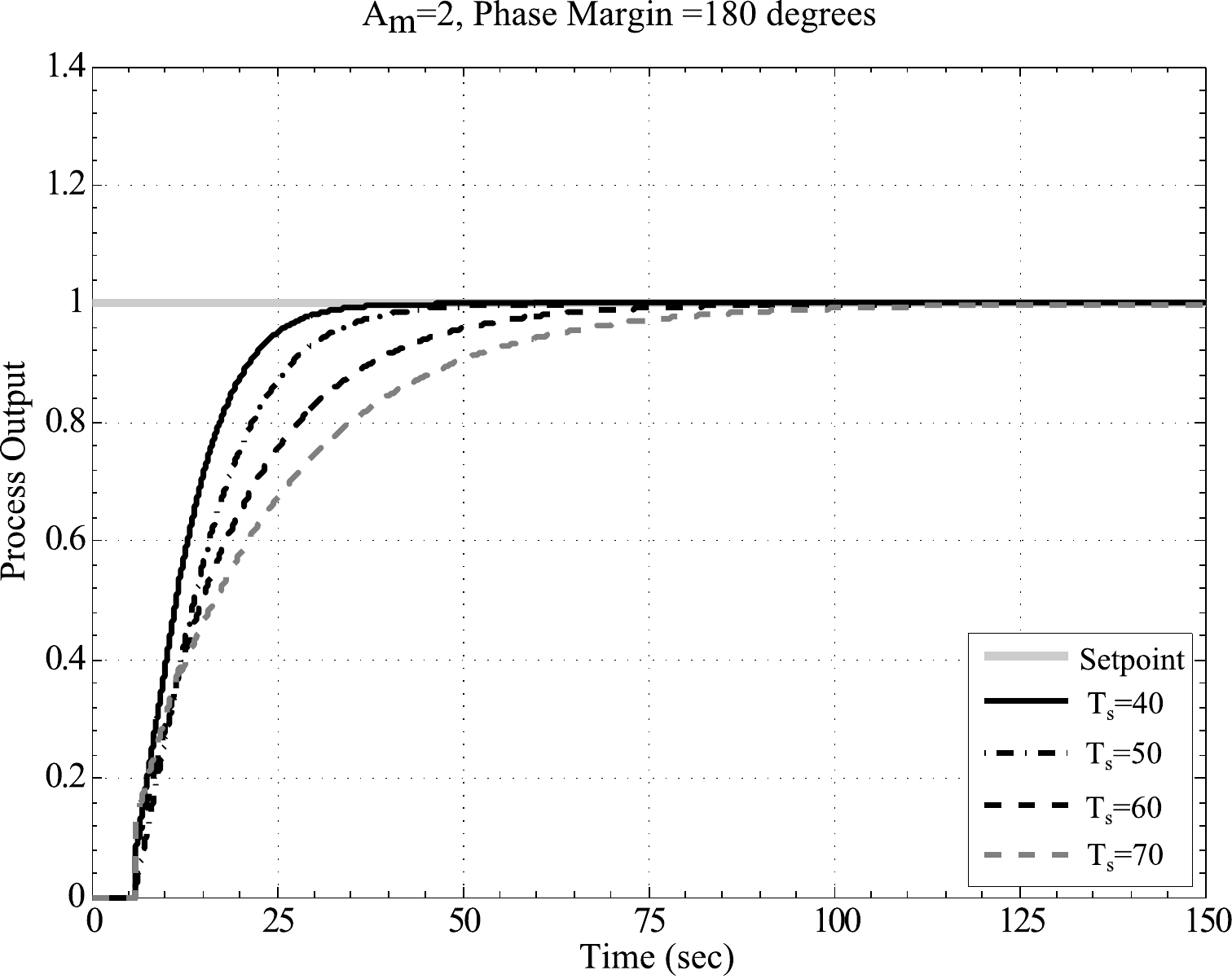}
\caption{Output responses when $A_m=2$ and $T_s=~40,~50,~60,~70.$}
\label{fig3}
\end{figure}
Above Fig. \ref{fig3} shows process output responses for the process given by equation (\ref{eq:IPDTsim}) when gain margin $A_m=2$, phase margin $\phi_m=180~ ^\circ$ and settling time $T_s~=~40,~50,~60,~70$. As the specified value of settling time increases, the speed of response gets slower. From Fig. \ref{fig3} it can be seen that the responses follow the specified settling time agreeably.
\begin{figure}[!htbp]
\centering
\includegraphics[width=0.6\linewidth]{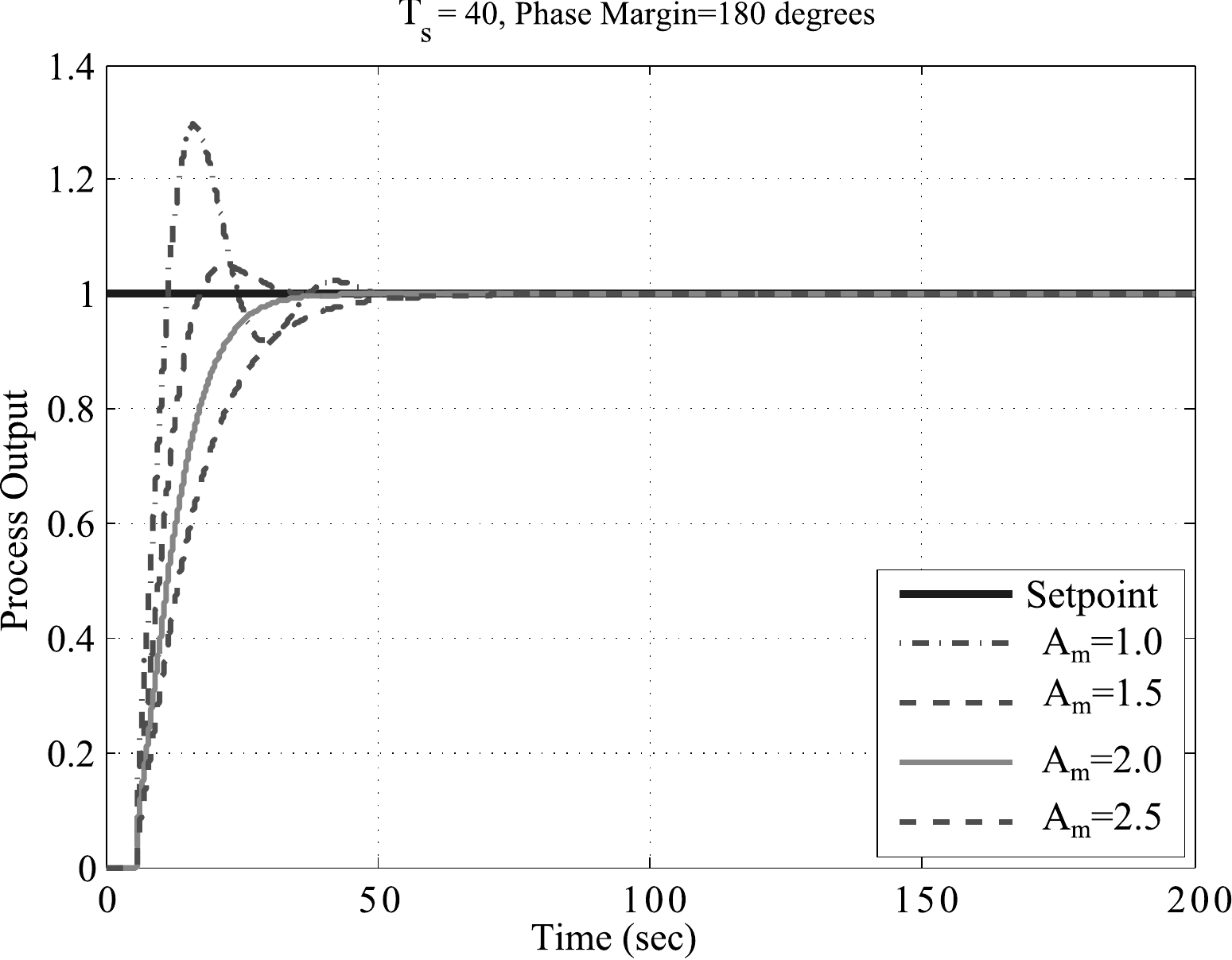}
\caption{Process output responses when  settling time $T_s=40$, phase margin $\phi_m=180 ^\circ$ and $A_m=~1,~1.5,~2,~2.5$.}
\label{fig4}
\end{figure}
Fig. \ref{fig4} shows the process output responses for the process given by equation (\ref{eq:IPDTsim}) when settling time $T_s~=~40$, phase margin $\phi_m=180 ^\circ$ and gain margin $A_m~=~1,~1.5,~2,~2.5$. Increase in gain margin specification increases the settling time, but decreases the oscillatory nature of the response i.e. it increases the damping coefficient of the overall system. A sufficiently small value of gain margin would result in an oscillatory response as in the case of $A_m=1$.
\section{Regulatory control using proposed PD controller}
\subsection{Use of disturbance observer with proposed PD controller}
The proposed controller being a PD controller, will fail to reject any permanent disturbances that act on the IPDT process. In order to make the proposed PD controller cope up with permanent load disturbances, a disturbance observer consisting of the IPDT process model $G_M(s)$ and a gain $Kc$ with the arrangement suggested in Fig. \ref{disobs} can be used. $D(s)$ is the unmeasured disturbance acting internally on the process and $\widehat{D}(s)$ is the approximation of the disturbance generated by the disturbance observer. Under influence of disturbance, the control signal applied to the IPDT process will be $U(s)~=~U_{PD}(s)-\widehat{D}(s)$. It is therefore essential that $\widehat{D}(s)$ closely follows $D(s)$.
\begin{figure}[!htbp]
\centering
\includegraphics[width=0.6\linewidth]{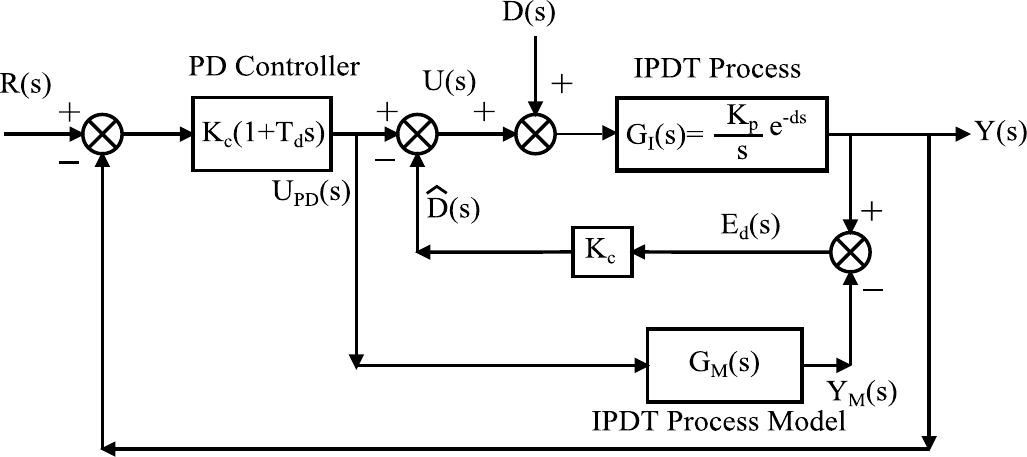}
\caption{Load disturbance rejection using PD controller and a disturbance observer}
\label{disobs}
\end{figure}

The behaviour of the disturbance observer can be described analytically by the equations that follow. With reference to Fig. \ref{disobs},
\begin{equation}
Y(s)=(U(s)+D(s)).G_I(s)
\end{equation}
\begin{equation}\label{eq:YMs}
Y_M(s)=U_{PD}(s).G_M(s)
\end{equation}
Also,
\begin{equation}\label{eq:Ys}
Y(s)=[U_{PD}(s)-\widehat{D}(s)+D(s)].G_M(s)
\end{equation}
and,
\begin{equation}\label{eq:DhatS}
\widehat{D}(s)=K_c[Y(s)-Y_M(s)]
\end{equation}
Using equations (\ref{eq:YMs}) and (\ref{eq:Ys}) in (\ref{eq:DhatS}),
\begin{equation}\label{eq:DHATS}
\widehat{D}(s)=K_c[G_I(s).(U_{PD}(s)-\widehat{D}(s)+D(s))-U_{PD}(s).G_M(s)]
\end{equation}
If perfect modeling is assumed,
\begin{equation}\label{eq:IdealModeling}
G_M(s)=G_I(s)
\end{equation}
Then, by rearranging above equation (\ref{eq:DHATS}), we have,
\begin{equation}\label{eq:disturbance}
\frac{\widehat{D}(s)}{D(s)}=\frac{G_M(s)K_c}{1+G_M(s)K_c}
\end{equation}
Equation (\ref{eq:disturbance}) represents the transfer function of the disturbance observer. From equation (\ref{eq:disturbance}) it should be clear that $\widehat{D}(s)$ would be present only if $D(s)$ is present. And, the gain $K_c$ present in the forward path of the block diagram shown in Fig. \ref{disobs} is necessary for quicker approximation of $D(s)$ and this has been determined after exhaustive simulations.
The alternative representation of the disturbance observer can be done as shown in Fig. \ref{alt_dist}.
\begin{figure}[!htbp]
\centering
\includegraphics[width=0.5\linewidth]{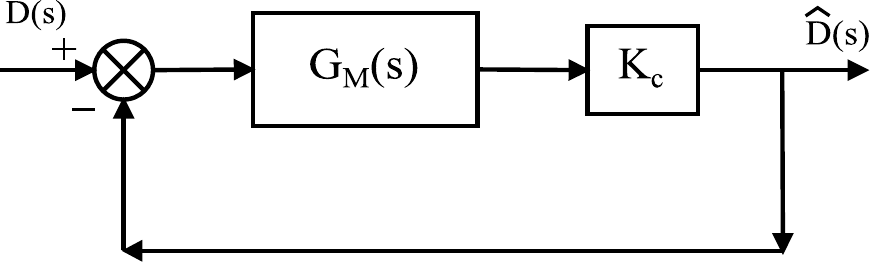}
\caption{Alternative representation of the disturbance  observer}
\label{alt_dist}
\end{figure}

\subsection{Regulatory responses}
Regulatory responses of the propose scheme and the methods under study, when the process is subjected to a step change disturbance are shown in Fig. \ref{reg_resp}.
\begin{figure}[!htbp]
\centering
\includegraphics[width=0.6\linewidth]{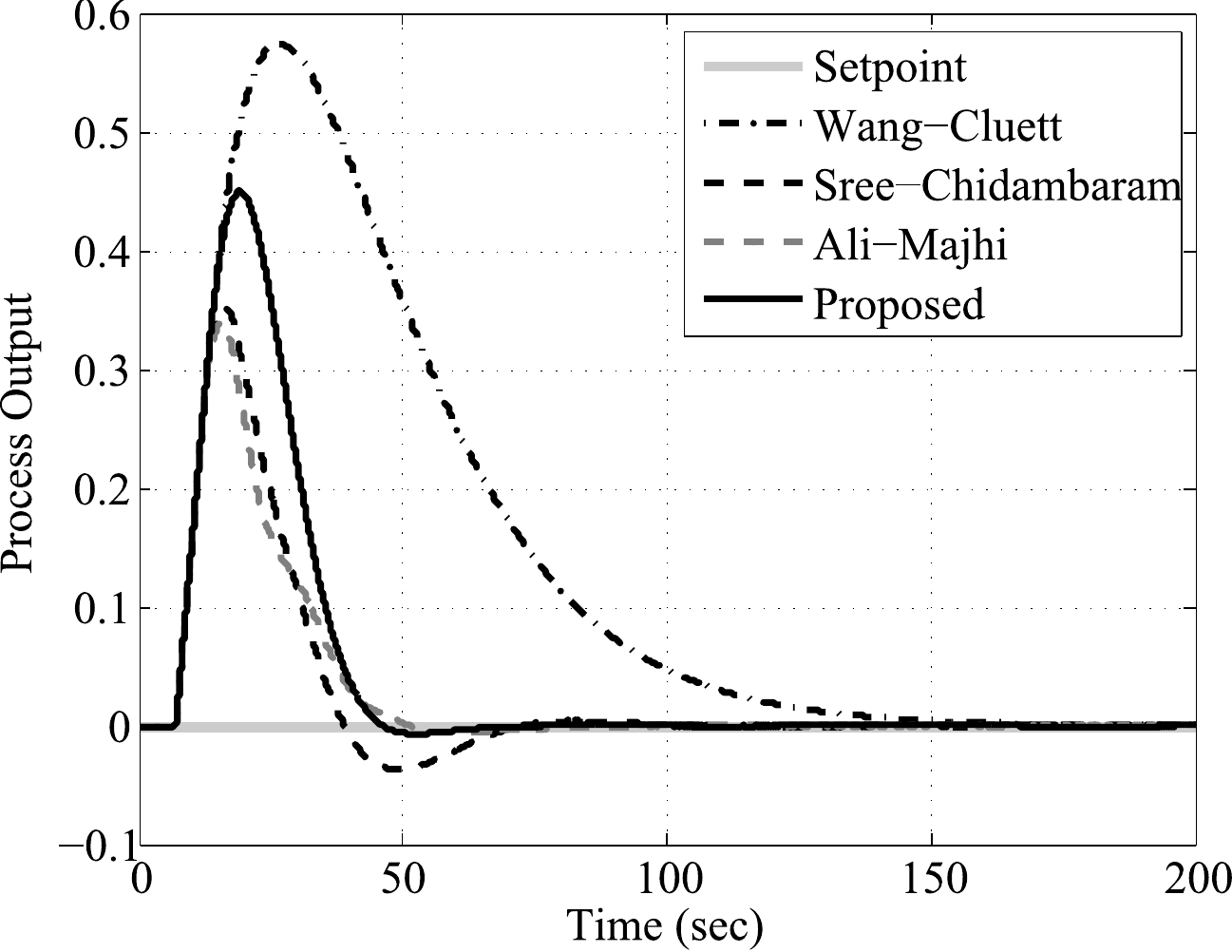}
\caption{Regulatory responses against a constant disturbance}
\label{reg_resp}
\end{figure}

The disturbance observer must approximate the disturbance affecting the process quickly for better regulatory response. The disturbance and its approximation generated by the disturbance observer are shown in Fig. \ref{dis_est}.
\begin{figure}[!htbp]
\centering
\includegraphics[width=0.6\linewidth]{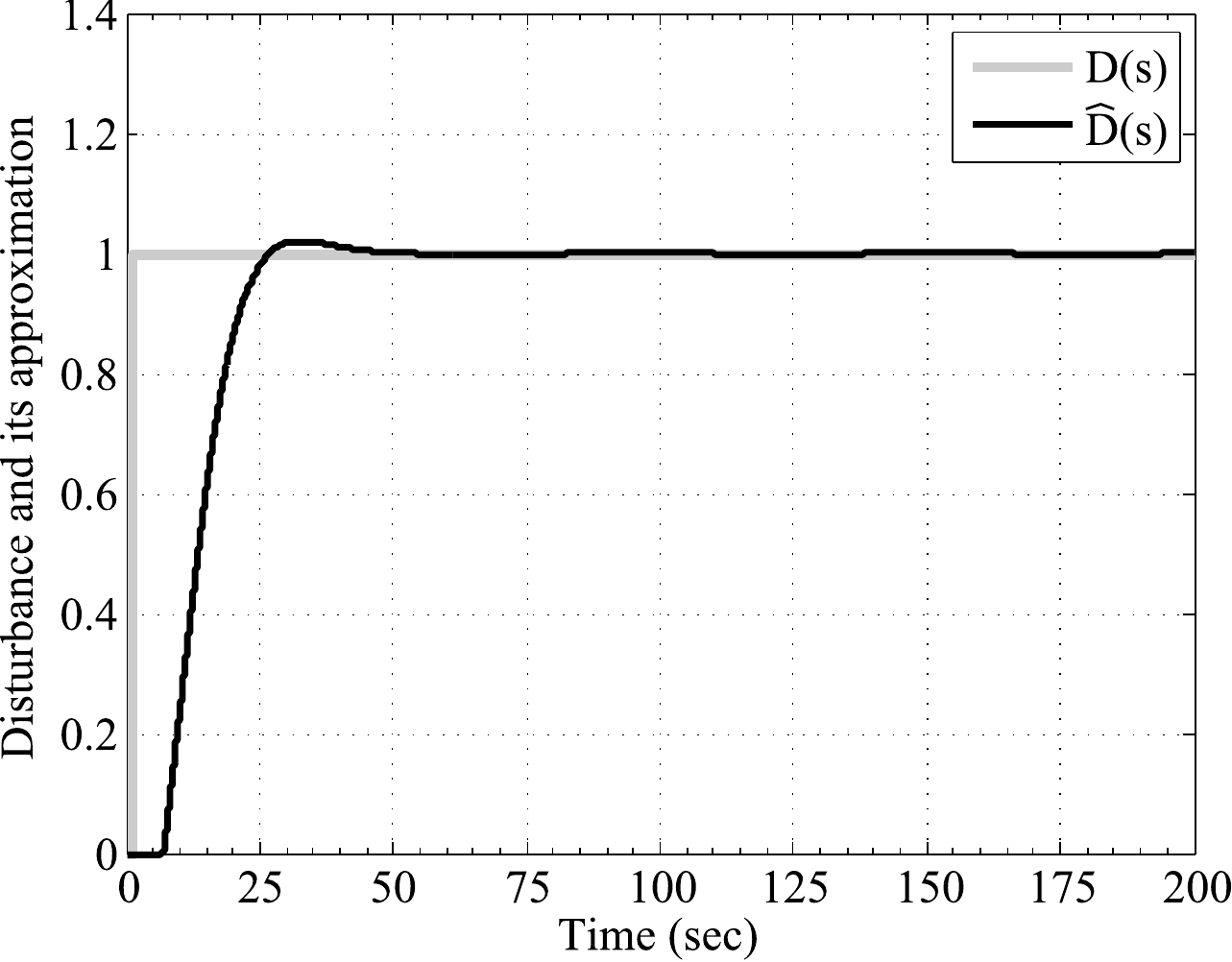}
\caption{Permanent  disturbance and its approximation by disturbance observer}
\label{dis_est}
\end{figure}
Table \ref{tab6} shows the comparison of proposed scheme and other tuning methods with respect to performances measures like ISE, IAE and control signal energy for regulatory responses obtained in Fig. \ref{reg_resp}.
\begin{table}[!htbp]
\caption{Comparison of regulatory responses}
\centering
\label{tab6}
\begin{tabular}{ c  c  c  c }
\hline
Parameters &\multicolumn{3}{c}{Regulatory} \\ \cline{2-4}
 &ISE &IAE &Control signal\\
 & & &energy\\
\hline
Wang-Cluett's tuning &20.41 &44.2 &210.9\\
\hline
Sree-Chidambaram's &1.4518 &6.526 &193.9\\
tuning & & &\\
\hline
Ali-Majhi's tuning &1.259 &5.842 &194.2\\
\hline
Proposed PD tuning &0.03 &1.002 &211.4\\
with disturbance observer & & &\\
\hline
\end{tabular}
\end{table}
\subsection{Servo plus regulatory responses}
Also, when the process is required to perform servo as well as regulatory control , the comparison of such responses obtained by using proposed scheme and other tuning methods is shown in Fig. \ref{fig9}.
\begin{figure}[!htbp]
\centering
\includegraphics[width=0.6\linewidth]{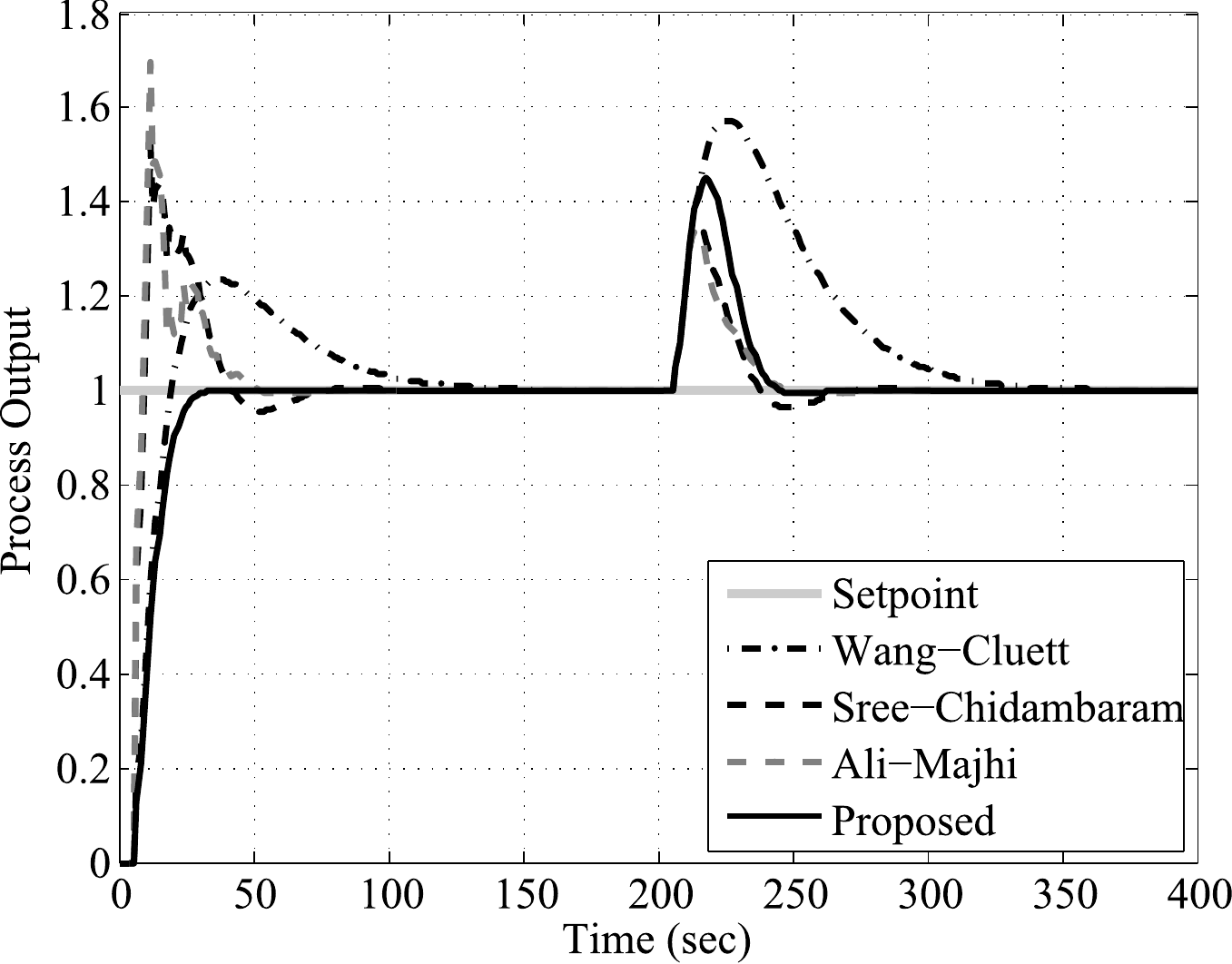}
\caption{Comparison of servo and regulatory responses when ${\rm SP}=1$ and constant disturbance enters process at $t=100~\rm{s}$  }
\label{fig9}
\end{figure}

\begin{table}[!htbp]
\caption{Comparison of servo and regulatory responses}
\centering
\label{table7}
\begin{tabular}{ c  c  c  c }
\hline
\hline
Parameters &\multicolumn{3}{c}{Servo and Regulatory} \\ \cline{2-4}
 &ISE &IAE &Control Signal\\
 & & &Energy\\
\hline
\hline
Wang-Cluett's tuning &21.83 &50.78 &425.6\\
\hline
Sree-Chidambaram's &10.23 &22.1 &4439\\
tuning & & &\\
\hline
Ali-Majhi's tuning &9.52 &19.49 &5743\\
\hline
Proposed PD tuning &12.69 &21.55 &289.7\\
with disturbance observer & & &\\
\hline
\hline
\end{tabular}
\end{table}
From comparisons in Table \ref{table7} the servo plus regulatory response of the proposed is acceptable when compared to other responses. At the same time, it can be seen that proposed control scheme is efficient in terms of control energy as reasonably less control effort is required as compared to other methods.
\section{Conclusion}
An approach for controller tuning comprising of both process transfer function based and frequency response based approaches was discussed for tuning PD controllers for integrating plus dead-time processes. The tuning rules can be easily computed by knowledge of process transfer function and intuitive specifications of settling time, gain and phase margins. Comparisons with Wang-Cluett's \cite{wang2}, Sree-Chidambaram's \cite{sree}, and  Ali-Majhi's \cite{ali} methods in terms of time domain specifications  show that smaller settling times for the integrating plus dead time processes can be obtained using proposed PD tuning rules. Control action taken by the proposed tuning rules is efficient as the control signal energy measure is appreciably below the control signal energy required by the other methods in order to control the process in case of servo or tracking control. The regulatory control has been made possible by using the proposed PD controller along with a disturbance observer. The regulatory responses of the proposed method  are fair and acceptable in comparison with the regulatory responses of other methods. Looking towards the combined servo and regulatory responses, the proposed scheme fares well over other methods in terms of control effort, as the disturbance observer contributes to the control action only when a disturbance acts upon the process, otherwise the control action remains the same as a PD controller's action.
\section{References}
\bibliographystyle{model3a-num-names}
\bibliography{FinRef}
\end{document}